\documentclass[letterpaper, 10 pt, conference]{ieeeconf}
\IEEEoverridecommandlockouts

\usepackage{cite}
\usepackage[utf8]{inputenc}
\usepackage{amsmath}
\usepackage{mathrsfs}
\usepackage{mathdots}
\usepackage{yhmath}
\usepackage{amssymb}
\usepackage{color}
\usepackage{dsfont}
\usepackage{comment}
\usepackage{float}
\usepackage{graphicx}
\usepackage{bernsteinStyle}
\usepackage{float}
\usepackage{tikz}
\usepackage{booktabs}
\usepackage{pgfplots}
\usepackage{array}
\usepackage{bm}
\usepackage{makecell}
\newcolumntype{C}[1]{>{\centering\let\newline\\\arraybackslash\hspace{0pt}}m{#1}}

\usetikzlibrary{arrows,shapes,positioning}

\usetikzlibrary{shapes,arrows,calc,positioning}
\tikzstyle{bigblock} = [draw, fill=blue!20, rectangle, 
    minimum height=6em, minimum width=8em]
\tikzstyle{medblock} = [draw, fill=blue!20, rectangle, 
    minimum height=4em, minimum width=4em]    
\tikzstyle{mux} = [draw, fill=black!20, rectangle, 
    minimum height=5em, minimum width=0.1em]    
\tikzstyle{smallblock} = [draw, fill=blue!20, rectangle, 
    minimum height=3em, minimum width=4em]
\tikzstyle{sum} = [draw, fill=blue!20, circle, node distance=1cm]
\tikzstyle{signal} = [coordinate]
\tikzstyle{pinstyle} = [pin edge={to-,thin,black}]
\tikzstyle{block} = [draw, fill=blue!20, rectangle, 
    minimum height=3em, minimum width=6em]
\tikzstyle{blockS} = [draw, fill=blue!20, rectangle, 
    minimum height=3em, minimum width=4em]  
\tikzstyle{sum} = [draw, fill=blue!20, circle, node distance=1.5cm]
\tikzstyle{gain} = [draw, fill=blue!20, regular polygon, regular polygon sides = 3, node distance=1.25cm, shape border rotate = -90]
\tikzstyle{mult} = [draw, fill=blue!20, circle, node distance=1.25cm]
\tikzstyle{input} = [coordinate]
\tikzstyle{output} = [coordinate]

\pgfplotsset{compat=1.16}

\title{A Data-Driven  Autopilot for Fixed-Wing Aircraft \\ Based on Model Predictive Control}
\author{\large Riley J. Richards, Juan A. Paredes, and Dennis S. Bernstein
\thanks{Riley J. Richards, Juan A. Paredes, and Dennis S. Bernstein are with the Department of Aerospace Engineering, University of Michigan, Ann Arbor, MI, USA. {\tt\small \{rileyric, jparedes, dsbaero\}@umich.edu}}
}

\begin{document}

\maketitle

\begin{abstract}
Autopilots for fixed-wing aircraft are typically designed based on linearized aerodynamic models consisting of stability and control derivatives obtained from wind-tunnel testing.
The resulting local controllers are then pieced together using gain scheduling.
For applications in which the aerodynamics are unmodeled, the present paper proposes an autopilot based on predictive cost adaptive control (PCAC).
As an indirect adaptive control extension of model predictive control, PCAC uses recursive least squares (RLS) with variable-rate forgetting for online, closed-loop system identification.
At each time step, RLS-based system identification updates the coefficients of an input-output model whose order is a hyperparameter specified by the user.
For MPC, the receding-horizon optimization can be performed by either the backward-propagating Riccati equation or quadratic programming.
The present paper investigates the performance of PCAC for fixed-wing aircraft without the use of any aerodynamic modeling or offline/prior data collection.
\end{abstract}

\section{Introduction}\label{sec:intro}

A fundamental necessity for autonomous atmospheric flight vehicles is a reliable autopilot for controlling the attitude and flight path. 
For a fixed-wing vehicle, stability and control derivatives are typically determined through wind-tunnel testing or computational modeling over a range of Mach number, angle of attack, and sideslip angle.
This modeling data is then used to develop an autopilot based on gain scheduling, feedback linearization, or dynamic inversion
\cite{tewari2011advanced,stengel2}.
In practice, however, the aerodynamics of an aircraft may be too expensive to model with high accuracy or may change due to atmospheric conditions, such as icing, as well as damage.
This possibility motivates the need to develop an adaptive autopilot that can learn and respond to unknown, changing conditions
\cite{hovakimyan2010,lavretsky2012robust,Ansari2017GTM}.

The present paper introduces a novel adaptive autopilot for fixed-wing aircraft based on model predictive control (MPC).
MPC is widely viewed as the most effective modern control technique, due to its ability to enforce state and control constraints in both linear and nonlinear systems through receding-horizon optimization 
\cite{kwon2006receding,camacho2013model,eren2017model,MPCbook}.

As its name suggests, MPC requires a model for optimization.
In the absence of a reliable model, data-driven techniques can be used to update the plant model during operation  
\cite{waardeinform,islam2021data,allgowerlearning2022}.
In effect, data-driven techniques perform closed-loop system identification, at least to a level of accuracy that is sufficient for the feedback controller to achieve desired closed-loop performance.
The interplay between system identification and control is a fundamental, longstanding problem in control theory
\cite{feldbaum,hjalmarsson1994,forssell1999closed,filatov,mesbahMPCdual2}.

To develop an adaptive autopilot for fixed-wing aircraft, the present paper focuses on predictive cost adaptive control (PCAC) \cite{islamPCAC,tamACC2021}, which is an indirect adaptive control extension of MPC.
For online, closed-loop system identification, PCAC uses recursive least squares (RLS) with variable-rate forgetting \cite{islam2019recursive,adamRLS2020,ankitRLS2020,bruceNandSRLS,mohseni2022recursive}.
At each time step, RLS-based system identification updates the coefficients of a SISO or MIMO input-output model, where the model order is a hyperparameter specified by the user.
For MPC, the receding-horizon optimization can be performed by either the backward-propagating Riccati equation \cite{kwonpearson,kwon2006receding} or quadratic programming.

The objective of the present paper is to investigate the performance of PCAC as a data-driven autopilot for an aircraft with unmodeled kinematics, dynamics, and aerodynamics.
In particular, PCAC is implemented as a cold-start indirect adaptive controller, where the plant model order is specified as a hyperparameter, but otherwise no plant model is assumed to be available.
The identified model updated by RLS is linear, and thus it is suitable for modeling the aircraft dynamics near trim.
In practice, an autopilot designed to operate over a wide range of flight conditions depends on gain scheduling of multiple linear controllers.
The goal of this study is to investigate, via numerical experiments, the viability and potential performance of PCAC under conditions of high uncertainty, in effect, no prior modeling information, without the need for gain scheduling.

For this study, we first consider scenarios involving the longitudinal and lateral dynamics of a linearized fixed-wing aircraft model provided by Athena Vortex Lattice (AVL). 
Next, we consider the longitudinal dynamics of a nonlinear fixed-wing  aircraft model provided in the MATLAB aerospace toolbox.



The contents of the paper are as follows.
Section \ref{sec:prob} introduces the predictive control problem.
Section \ref{sec:PCAC} reviews the PCAC formulation and algorithm.
Section \ref{sec:PCAC_exam1} applies PCAC to a linear 6DOF aircraft model for longitudinal and lateral control, and 
Section \ref{sec:PCAC_exam2} applies PCAC to a nonlinear 3DOF aircraft model for longitudinal control.
Finally, Section \ref{sec:conclusions} presents conclusions and future research directions.

{\bf Notation:}
$\bfz\in\BBC$ denotes the Z-transform variable.
$x_{(i)}$ denotes the $i$th component of $x\in\BBR^n.$
${\rm sprad}(A)$ denotes the spectral radius of $A\in\BBR^{n\times n}.$
The symmetric matrix $P\in\BBR^{n\times n}$ is positive semidefinite (resp., positive definite) if all of its eigenvalues are nonnegative (resp., positive).
$\vek X\in\BBR^{nm}$ denotes the vector formed by stacking the columns of $X\in\BBR^{n\times m}$, and $\otimes$ denotes the Kronecker product.
$I_n$ is the $n \times n$ identity matrix, and $0_{n\times m}$ is the $n\times m$ zeros matrix and $\mathds{1}_{n\times m}$ is the $n\times m$ ones matrix.

\section{Statement of the Control Problem}\label{sec:prob}
\begin{figure} [h!]
    \centering
    \vspace{0.25em}
    \resizebox{0.8\columnwidth}{!}{%
    \begin{tikzpicture}[>={stealth'}, line width = 0.25mm]

    \node [input, name=ref]{};
    \node [smallblock, rounded corners, right = 0.5cm of ref , minimum height = 1.2cm, minimum width = 1cm] (controller) {$G_{\rmc,k}$};
    \node[smallblock, rounded corners, right = 1.2cm of controller, minimum height = 0.6cm, minimum width = 0.5cm](sat_Blk){$\sigma$};
    \node[smallblock, rounded corners, right = 0.5cm of sat_Blk, minimum height = 0.6cm, minimum width = 0.5cm](rate_Blk){$\gamma_k$};
    \node [smallblock, rounded corners, right = 0.75cm of rate_Blk, minimum height = 0.6cm , minimum width = 0.5cm] (DA) {ZOH};
    \node[smallblock, rounded corners, below = 0.18cm of controller, minimum height = 0.6cm, minimum width = 0.5cm](ytGain){$C_\rmt$};
    
    \node [smallblock, fill=green!20, rounded corners, right = 0.75cm of DA, minimum height = 0.6cm , minimum width = 1cm] (system) {$\SM$};
    \node [output, right = 0.5cm of system] (output) {};
    \node [input, below = 1.3cm of system] (midpoint) {};
    
    \draw [->] (controller) -- node [above] {$u_{{\rm req},k}$} (sat_Blk);\
    \draw [->] (sat_Blk) -- (rate_Blk);
    \draw [->] (rate_Blk) -- node [above] {$u_k$} (DA);\
    \draw [->] (DA) -- node [above] {$u (t)$} (system);
    
    \node[circle,draw=black, fill=white, inner sep=0pt,minimum size=3pt] (rc11) at ([xshift=-2.5cm]midpoint) {};
    \node[circle,draw=black, fill=white, inner sep=0pt,minimum size=3pt] (rc21) at ([xshift=-2.8cm]midpoint) {};
    \draw [-] (rc21.north east) --node[below,yshift=.55cm]{$T_\rms$} ([xshift=.3cm,yshift=.15cm]rc21.north east) {};
    
    \draw [->] (system) -- node [name=y, near end]{} node [very near end, above] {$y (t)$}(output);

    \draw [-] (y.west) |- (midpoint);
    \draw [-] (midpoint) -| (rc11.east);
    \draw [->] (rc21) -| ([xshift = -1.25cm, yshift = 0.3cm]controller.west) -- node [above, xshift = 0.25cm] {$y_k$} ([yshift = 0.3cm]controller.west);

    \draw[->] ([yshift = 0.5cm]controller.north) -- node[xshift = 0.25cm,yshift=0.05cm]{$r_k$} (controller.north);
    \draw[->] (ytGain.west) -| ([xshift = -0.75cm, yshift = -0.3cm]controller.west) --  node [above] {$y_{\rmt,k}$} ([yshift = -0.3cm]controller.west);

    \draw[->] ([xshift = -5.2cm]midpoint) |- (ytGain.east);
    
    \end{tikzpicture}
    }  
    \caption{Sampled-data implementation of predictive controller for control of continuous-time system $\SM.$
    All sample-and-hold operations are synchronous.
    The predictive controller $G_{\rmc,k}$ generates the requested discrete-time control $u_{{\rm req},k}\in\BBR^m$ at each step $k$.
    The implemented discrete-time control is $u_k=\gamma_k(\sigma(u_{{\rm req},k}))$, where $\sigma\colon\BBR^m\to\BBR^m$ represents control-magnitude saturation and $\gamma_k \colon\BBR^m\to\BBR^m$ represents move-size saturation.
    The resulting continuous-time control $u(t)$ is generated by applying a zero-order-hold operation to $u_k$.
    For this work, $\SM$ represents a fixed-wing aircraft simulation model.}
    \vspace{-0.1in}
    \label{fig:PC_CT_blk_diag}
\end{figure}
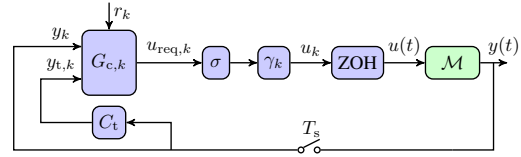

To reflect the practical implementation of digital controllers for physical systems, we consider continuous-time dynamics under sampled-data control using discrete-time predictive controllers.
In particular, we consider the control architecture shown in Figure \ref{fig:PC_CT_blk_diag}, where $\SM$ is the target continuous-time system, $t\ge 0$, $u(t)\in\BBR^m$ is the control, and $y(t)\in\BBR^p$ is the output of $\SM,$ which is sampled to produce the measurement $y_k \in \BBR^p,$ which, for all $k\ge0,$ is given by
\begin{equation}
    y_k \isdef  y(k T_\rms),
\end{equation}
where  $T_\rms>0$ is the sample time.

The predictive controller, which is updated at each step $k,$ is denoted by $G_{\rmc,k}$.
For all $k\ge0,$ let $y_{\rmt,k} \isdef C_{\rmt, k} y_k \in \BBR^{p_\rmt}$ be the command following output, where $C_{\rmt, k} \in \BBR^{p_\rmt \times p},$ and let $r_k \in \BBR^{p_\rmt}$ be the  command.
The inputs to $G_{\rmc,k}$ are $r_k,$ $y_k,$ and $y_{\rmt, k},$ and the output is the requested discrete-time control $u_{{\rm req},k}\in\BBR^m.$
The predictive controller uses $y_k$ for system identification, and $r_k$ and $y_{\rmt, k}$ trajectory command following.
Since the response of a real actuator is subjected to hardware constraints, the implemented discrete-time control is 
\begin{equation}
 u_k\isdef \gamma_k(\sigma(u_{{\rm req},k})),  
 \label{eq:ukksat}
\end{equation}
where $\sigma\colon\BBR^m\to\BBR^m$ is the control magnitude saturation function  
\begin{equation}
    \sigma(u) \isdef  \matl \bar{\sigma}(u_{(1)}) \\ \vdots \\ \bar{\sigma} (u_{(m)})\matr,
    \label{eq:Hsat1}
\end{equation}
where $\bar\sigma\colon \BBR\to\BBR$ is defined by 
\begin{equation}
    \bar\sigma(u)\isdef \begin{cases} u_{\max},&u> u_{\max},\\
    u,& u_{\min}\le u \le u_{\max},\\
    u_{\min}, & u<u_{\min},\end{cases}
\end{equation}
and $u_{\min},u_{\max}\in\BBR$ are the lower and upper magnitude saturation levels, respectively,
and $\gamma_k\colon\BBR^m\to\BBR^m$ is the control rate (move-size) saturation function  
\begin{equation}
    \gamma_k(u_k) \isdef  \matl \bar{\gamma}_k(u_{k (1)}) \\ \vdots \\ \bar{\gamma}_k (u_{k (m)})\matr,
    \label{eq:Hsat2}
\end{equation}
where $\bar{\gamma}_k \colon \BBR\to\BBR$ is defined by 
\small
\begin{equation}
    \bar{\gamma}_k (u_k)\isdef \begin{cases} u_{k-1} + \Delta u_{\max},& u_k - u_{k-1} > \Delta u_{\max},\\
    u_k,& \Delta u_{\min}\le u_k - u_{k-1} \le \Delta u_{\max},\\
    u_{k-1} + \Delta u_{\min}, & u_k - u_{k-1}< \Delta u_{\min},\end{cases}
\end{equation}
\normalsize
and $\Delta u_{\min}, \Delta u_{\max}\in\BBR$ are the lower and upper move-size saturation levels, respectively.
%
%
The continuous-time control signal $u(t)$ applied to the structure is generated by applying a zero-order-hold operation to $u_k,$ that is,
for all $k\ge0,$ and, for all $t\in[kT_\rms, (k+1) T_\rms),$ 
\begin{equation}
    u(t) = u_k.
\end{equation}
%
%
The objective of the predictive controller is to yield an input signal that minimizes the norm of the difference between the command following output and the  command, that is, yield $u_k$ such that $\sum_{k = 0}^\infty ||y_{\rmt,k} - r_k||_2$ is minimized.

\section{Review of Predictive Cost Adaptive Control}\label{sec:PCAC}

PCAC combines online identification  with output-feedback MPC. 
The PCAC algorithm is presented in this section.
Subsection \ref{subsec:ID} describes the technique used for online identification, namely, RLS with variable-rate forgetting based on the F-test \cite{mohseni2022recursive}.
Subsection \ref{subsec:bocf} presents the block observable canonical form (BOCF), which is used to represent the input-output dynamics model as a state space model whose state is given explicitly in terms of inputs, outputs, and model-coefficient estimates.
Subsection \ref{subsec:MPC} reviews the MPC technique for receding-horizon optimization.
%

\subsection{Online Identification Using Recursive Least Squares with Variable-Rate Forgetting
Based on the F-Test} \label{subsec:ID}

Let $\hat n\ge 0$ and, for all $k\ge 0,$ let $F_{\rmm,1,k},\hdots, F_{\rmm,\hat n,k}\in\BBR^{p\times p}$ and $G_{\rmm,1,k},\hdots, G_{\rmm,\hat n,k}\in\BBR^{p\times m}$ be the coefficient matrices to be estimated using RLS.
Furthermore, let $\hat y_k\in\BBR^p$ be an estimate of $y_k$ defined  by
\begin{equation}
\hat y_k\isdef -\sum_{i=1}^{\hat n}  F_{\rmm,i,k}   y_{k-i} + \sum_{i=1}^{\hat n} {G}_{\rmm,i,k} u_{k-i},
\label{eq:yhat}
\end{equation}
where  
\begin{gather}
   y_{-\hat n}=\cdots= y_{-1}=0,\\ u_{-\hat n}=\cdots=u_{-1}=0. 
\end{gather}
Using the identity ${\rm vec} (XY) = (Y^\rmT \otimes I) {\rm vec} X,$ it follows from  \eqref{eq:yhat} that, for all $k\ge 0,$ 
\begin{equation}
    \hat y_k = \phi_k \theta_k,
    \label{eq:yhat_phi}
\end{equation}
where
\begin{align}
     %
     \theta_k \isdef & \ \matl \theta_{F_\rmm, k}^{\rm T} & \theta_{G_\rmm, k}^{\rm T} \matr^{\rm T} \in\BBR^{\hat np(m+p)},\\
     \theta_{F_\rmm, k} \isdef & \ {\rm vec}\matl  F_{\rmm,1,k}&\cdots& F_{\rmm,\hat n,k} \matr \in\BBR^{\hat n p^2}, \\
     \theta_{G_\rmm, k} \isdef & \ {\rm vec}\matl  G_{\rmm,1,k}&\cdots& G_{\rmm,\hat n,k} \matr \in\BBR^{\hat n pm}, \\
     \phi_k \isdef & \matl -  y_{k-1}^{\rm T}&\cdots&- y_{k-\hat n}^{\rm T}&u_{k-1}^{\rm T}&\cdots& u_{k-\hat n}^{\rm T}\matr\otimes I_p \nn \\ &\in\BBR^{p\times \hat np(m+p)}.
\label{eq:phi_kkka}
 \end{align}

To determine the update equations for $\theta_k$, for all $k\ge 0$, define $e_k\colon\BBR^{\hat np(m+p)}\to\BBR^p$ by
\begin{equation}
    e_k(\bar \theta) \isdef y_k - \phi_k \bar \theta,
    \label{eq:ekkea}
\end{equation}
where $\bar \theta\in\BBR^{\hat np(m+p)}.$ 
Using  \eqref{eq:yhat_phi}, the \textit{identification error} at step $k$ is defined by
\begin{equation}
 e_k(\theta_k)= y_k-\hat y_k.  
\end{equation}
For all $k\ge 0$, the RLS cumulative cost $J_k\colon\BBR^{\hat np(m+p)}\to[0,\infty)$ is defined by \cite{islam2019recursive}
\begin{equation}
J_k(\bar \theta) \isdef \sum_{i=0}^k \frac{\rho_i}{\rho_k} e_i^{\rm T}(\bar \theta) e_i(\bar \theta) + \frac{1}{\rho_k} (\bar\theta -\theta_0)^{\rm T} \psi_0^{-1}(\bar\theta-\theta_0),
\label{Jkdefn}
\end{equation}
where $\psi_0\in\BBR^{\hat np(m+p)\times \hat np(m+p)}$ is  positive definite, $\theta_0\in\BBR^{\hat n p(m+p)}$ is the initial estimate of the coefficient vector, and, for all $i\ge 0,$
\begin{equation}
  \rho_i \isdef \prod_{j=0}^i \lambda_j^{-1}.  
\end{equation}
For all $j\ge 0$, the parameter $\lambda_j\in(0,1]$ is the forgetting factor defined by $\lambda_j\isdef\beta_j^{-1}$, where
\begin{equation}
 \beta_j \isdef \begin{cases}
     1, & j<\tau_\rmd,\\
     1 + \eta \bar{\beta}_j,& j\ge \tau_\rmd,
 \end{cases}    
\end{equation}
\footnotesize
\begin{equation}
    \bar{\beta}_j \isdef g(e_{j-\tau_\rmd}(\theta_{j-\tau_\rmd}),\hdots,e_j(\theta_j)) \textbf{1}\big(g(e_{j-\tau_\rmd}(\theta_{j-\tau_\rmd}),\hdots,e_j(\theta_j))\big),
\end{equation}
\normalsize
and  $\tau_\rmd> p$, $\eta>0$,  $\textbf{1}\colon \BBR\to\{0,1\}$ is the unit step function, and $g$ is a function of past RLS identification errors which is given by (10) in \cite{mohseni2022recursive} in the case where $p = 1$ and (13) in \cite{mohseni2022recursive} in the case where $p > 1.$
Note that $g$ includes forgetting terms based on the inverse cumulative distribution function of the F-distribution and depends on $\tau_\rmd,$ $\tau_\rmn\in[p,\tau_\rmd),$ and \textit{significance level} $\alpha_F\in (0,1].$

Finally, for all $k\ge0$, the unique global minimizer of $J_k$ is given by \cite{islam2019recursive}
\begin{equation}
    \theta_{k+1} = \theta_k +\psi_{k+1} \phi_k^{\rm T} (y_k - \phi_k \theta_k),
\end{equation}
where 
\begin{align}
  \psi_{k+1} &\isdef  \beta_k \psi_k - \beta_k \psi_k \phi_k^{\rm T} (\tfrac{1}{\beta_k}I_p + \phi_k  \psi_k \phi_k^{\rm T})^{-1} \phi_k  \psi_k,
\end{align}
and $\psi_0$ is the performance-regularization weighting in \eqref{Jkdefn}.
Additional details concerning RLS with forgetting based on the F-distribution are given in \cite{mohseni2022recursive}.


\subsection{Input-Output Model and the Block Observable Canonical Form} \label{subsec:bocf}

Considering the estimate $\hat y_k$ of $y_k$ given by \eqref{eq:yhat}, it follows that, for all $k\ge0,$
\begin{equation}
y_{k} \approx -\sum_{i=1}^{\hat n}  F_{\rmm,i,k}   y_{k-i} + \sum_{i=1}^{\hat n} {G}_{\rmm,i,k} u_{k-i}.
\label{eq:ykapp}
\end{equation}
Viewing \eqref{eq:ykapp} as an equality, it follows that, for all $k\ge 0,$ the BOCF state-space realization of \eqref{eq:ykapp} is given by  \cite{polderman1989state}
\begin{align}
x_{\rmm,k+1} &=   A_{\rmm,k}  x_{\rmm,k} +  B_{\rmm,k} u_k,\label{eq:xmssAB}\\
y_{k} &=  C_\rmm   x_{\rmm, k}, \label{eq:yhatCxm}
\end{align}
where
\begin{gather}
 A_{\rmm,k} \isdef \matl - F_{\rmm,1,k+1} & I_p & \cdots & \cdots & 0_{p\times p}\\
- F_{\rmm,2,k+1} & 0_{p\times p} & \ddots & & \vdots\\
\vdots & {\vdots} & \ddots & \ddots & 0_{p\times p} \\
\vdots & \vdots &  & \ddots & I_p\\
- F_{\rmm,\hat n,k+1} & 0_{p\times p} & \cdots &\cdots & 0_{p\times p}
\matr\in\BBR^{\hat np\times \hat n p},\label{eq:ABBA_1}\\
B_{\rmm,k}\isdef \matl  G_{\rmm,1,k+1} \\
 G_{\rmm,2,k+1}\\
\vdots\\
 G_{\rmm,\hat n,k+1}
\matr \in\BBR^{\hat n p \times m},\label{eq:ABBA_2}\\
  C_\rmm\isdef \matl I_p & 0_{p\times p} & \cdots & 0_{p\times p} \matr\in\BBR^{p\times \hat n p},
  \label{eq:cmmx}
\end{gather}
and 
\begin{equation}
    x_{\rmm,k} \isdef \matl x_{\rmm,k(1)}\\\vdots\\ x_{\rmm,k(\hat n)}\matr \in\BBR^{\hat n p},
    \label{eq:xmkkx}
\end{equation}
where
\vspace{-0.1in}
\begin{align}
 x_{\rmm,k(1)} \isdef  y_{k},
\label{eq:x1kk}
\end{align}
and, for all $j=2,\ldots,\hat n,$
\begin{align}
 x_{\rmm,k(j)} \isdef & -\sum_{i=1}^{\hat n -j +1}  F_{\rmm,i+j-1,k+1}  y_{k-i} \nn \\
 &+ \sum_{i=1}^{\hat n -j+1}  G_{\rmm,i+j-1,k+1} u_{k-i}.
 \label{eq:xnkk}
\end{align}

\noindent Note that multiplying both sides of \eqref{eq:xmssAB} by $C_\rmm$ and using \eqref{eq:yhatCxm}--\eqref{eq:xnkk} implies that, for all $k\ge 0,$
\begin{align}
     y_{k+1}&=C_\rmm x_{\rmm,k+1}\nn\\
     &= C_\rmm(A_{\rmm,k}  x_{\rmm,k} +  B_{\rmm,k} u_k)\nn\\
    &=-F_{\rmm,1,k+1}  x_{\rmm,k(1)} + x_{\rmm,k(2)} +G_{\rmm,1,k+1} u_k \nn\\
    &=-F_{\rmm,1,k+1}  y_{k}  -\sum_{i=1}^{\hat n -1}  F_{\rmm,i+1,k+1}  y_{k-i} \nn\\
    &+ \sum_{i=1}^{\hat n -1}  G_{\rmm,i+1,k+1} u_{k-i}+G_{\rmm,1,k+1} u_k \nn\\
    &=  -\sum_{i=1}^{\hat n }  F_{\rmm,i,k+1}  y_{k+1-i} + \sum_{i=1}^{\hat n }  G_{\rmm,i,k+1} u_{k+1-i},
\end{align}
which is approximately equivalent to \eqref{eq:ykapp} with $k$ in \eqref{eq:ykapp} replaced by $k+1$.


\subsection{Model Predictive Control (MPC)} \label{subsec:MPC}

Let $\ell\ge 1$ be the horizon and, for all $k\ge0$ and all $i=1,\ldots,\ell,$ let $x_{\rmm,k|i}\in\BBR^{\hat n p}$ be the $i$-step predicted state, $y_{\rmm,k|i}\in\BBR^p$ be the $i$-step predicted output, and $u_{k|i}\in\BBR^m$ be the $i$-step predicted control.
Then, the $\ell$-step predicted output of \eqref{eq:yhatCxm} for a sequence of $\ell$ future controls is given by
\begin{align}
    Y_{\ell, k|1} = \Gamma_{\ell, k} x_{\rmm,k|1} + T_{\ell, k} U_{\ell, k|1},
\end{align}
where
\begin{align}
    Y_{\ell, k|1} \isdef \begin{bmatrix} y_{\rmm, k|1} \\ \vdots \\ y_{\rmm,k|\ell}  \end{bmatrix} \in \BBR^{\ell p}, \quad
    U_{\ell, k|1} \isdef \begin{bmatrix} u_{k|1} \\ \vdots \\ u_{k|\ell}  \end{bmatrix} \in \BBR^{\ell m}, 
\end{align}
%
%
\begin{equation}
    \Gamma_{\ell, k} \isdef \begin{bmatrix} C_\rmm^{\rm T} & (C_\rmm A_{\rmm, k})^{\rm T} & \cdots & (C_\rmm A_{\rmm, k}^{\ell - 1})^{\rm T} \end{bmatrix}^{\rm T} \in \BBR^{\ell p \times \hat{n} p},
\end{equation}
\small
\begin{align}
    T_{\ell, k} &\isdef
    \matl
    0_{p \times m} & \cdots & \cdots & \cdots & \cdots & 0_{p\times m} \\
    H_{k,1} & 0_{p \times m} & \cdots & \cdots & \cdots & 0_{p\times m} \\
    H_{k,2} & H_{k,1} & 0_{p \times m} & \cdots  & \cdots & 0_{p\times m} \\ 
    H_{k,3} & H_{k,2} & H_{k,1} & 0_{p \times m}  & \cdots & 0_{p\times m} \\
    \vdots & \vdots & \vdots & \ddots & \ddots & 0_{p\times m} \\
    H_{k,\ell-1} & H_{k,\ell-2} & H_{k,\ell-3} & \cdots & H_{k,1} & 0_{p \times m}
    \matr \nn \\
    & \in \BBR^{\ell p \times \ell m},
\end{align}
\normalsize
where $H_{k,i} \isdef C_\rmm A_{\rmm,k}^{i-1} B_{\rmm,k} \in \BBR^{p \times m}$ for all $i=1,\ldots,\ell-1.$

Let $\mathcal{R}_{\ell,k} \isdef \begin{bmatrix} r_{k+1}^{\rm T} \ \cdots \ r_{k+\ell}^{\rm T} \end{bmatrix}^{\rm T} \in \BBR^{\ell p_\rmt}$ be a vector composed of $\ell$ future commands,
let $y_{\rmm,\rmt,k|i} \isdef C_\rmt y_{\rmm,k|i} \in\BBR^{p_\rmt}$ be the $i$-step predicted command-following output,
let $Y_{\rmt, \ell , k|1} \isdef \matl y_{\rmm, \rmt, k|1}^{\rm T} & \cdots & y_{\rmm, \rmt, k|\ell}^{\rm T} \matr^{\rm T} = C_{\rmt,\ell} Y_{\ell, k|1}\in \BBR^{\ell p_\rmt},$ where $C_{\rmt,\ell} \isdef I_\ell \otimes C_\rmt \in \BBR^{\ell p_\rmt \times \ell p},$
and define
\begin{align}
    \Delta U_{\ell, k|1} \isdef 
    \matl
     u_{k|1} - u_k   \\  u_{k|2} - u_{k|1}   \\ \vdots \\  u_{k|\ell} - u_{k|\ell - 1}   \matr \in \BBR^{\ell m }.
\end{align}
Then, the receding horizon optimization problem is given by
\begin{align}
    \min_{U_{\ell,k|1}} \left(Y_{\rmt, \ell, k|1} - \mathcal{R}_{\ell,k}\right)^{\rm T} & Q \left(Y_{\rmt, \ell, k|1} - \mathcal{R}_{\ell,k}\right) \nn \\ &\quad + \Delta U_{\ell, k|1}^{\rm T} R \Delta U_{\ell, k|1},
\end{align}
subject to
\begin{align}
U_{\min} \leq U_{\ell, k|1} &\leq U_{\max} \\
\Delta U_{\min} \leq \Delta U_{\ell, k|1} &\leq \Delta U_{\max},
\end{align}
where $Q \triangleq \matl \bar{Q} & 0_{p_\text{t} \times p_\text{t}} \\ 0_{p_\text{t} \times p_\text{t}} & \bar{P} \matr \in \mathbb{R}^{\ell p_\text{t} \times \ell p_\text{t}}$ is the positive-definite output weighting, $\bar{Q} \in \mathbb{R}^{(\ell -1) p_\text{t} \times (\ell -1) p_\text{t}}$ is the positive-definite cost-to-go output weighting, $\bar{P} \in \mathbb{R}^{p_\text{t} \times p_\text{t}}$ is the positive-definite terminal output weighting,
$R \in \BBR^{\ell m \times \ell m}$ is the positive definite control move-size weight, $U_{\min} \isdef 1_{\ell } \otimes u_{\min} \in \BBR^{\ell m}$, $U_{\max} \isdef 1_{\ell } \otimes u_{\max} \in \BBR^{\ell m}$, $\Delta U_{\min} \isdef 1_{\ell } \otimes \Delta u_{\min} \in \BBR^{\ell m}$, and $\Delta U_{\max} \isdef 1_{\ell } \otimes \Delta u_{\max} \in \BBR^{\ell m}$.

In summary, at each time step,  online identification is performed to find input-output model coefficients $\theta_{k+1}$, which are then used to create a state space realization $\left(A_{\rmm, k}, B_{\rmm, k}, C_\rmm \right)$.
Then, the state-space realization is used in a receding horizon optimization problem to solve for the $\ell$-step  controls $U_{\ell, k|1}.$
The control input for the next step is then given by $u_{k|1}$, and the rest of the components of $U_{\ell, k|1}$ are discarded.

\section{Adaptive Control of a Linearized 6DOF Aircraft Model} \label{sec:PCAC_exam1}

In this section, a linearized 6DOF fixed-wing RC aircraft model from Athena Vortex Lattice (AVL) is considered for both longitudinal and lateral control. 
The first example uses altitude, elevation angle, and bank angle measurements for altitude and bank-angle command following using the elevator and aileron. 
The second example uses altitude, elevation, bank, and azimuth measurements for altitude and azimuth-angle command following using all three control surfaces.
Note that all numerical examples in this section use a sampling rate of $T_s = 0.05$ sec/step and the aircraft model is initialized at an altitude $h_0 = 500$ m and airspeed $V_0 = 10$ m/s with all other initial model parameters set to 0.

\subsection{Altitude and Bank-Angle Command Following} \label{ABT}

\begin{figure} [h!]
    \centering
    \resizebox{\columnwidth}{!}{%
    \includegraphics{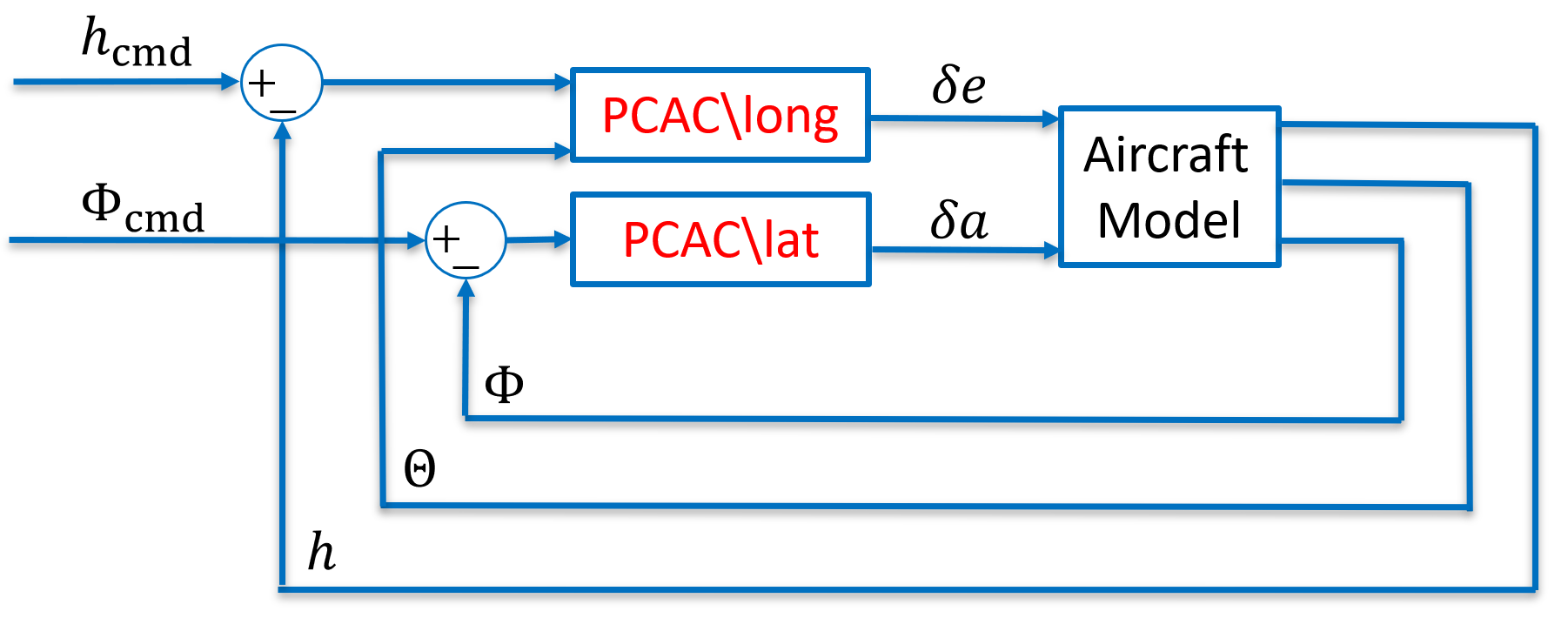}
    }
    \caption{Command-following block diagram for altitude $h$ and bank angle $\Phi$. 
    This architecture uses two PCAC controllers, namely, PCAC/long for the longitudinal dynamics and PCAC/lat for the lateral dynamics.
    Note that PCAC/long uses the elevation angle $\Theta$ for system identification.}
    \label{Case1BD}
    \vspace{-0.1in}
\end{figure}

For this example, two PCAC loops are used to follow altitude and bank-angle commands.
The longitudinal loop uses the altitude $h$ and elevation angle $\Theta$ to identify the longitudinal dynamics and to specify the elevator deflection $\delta e$. 
%
The lateral loop uses the bank angle $\Phi$ to identify the lateral dynamics and to specify the aileron deflection $\delta a$.
This control architecture is illustrated in Figure \ref{Case1BD}.

\begin{figure} [h!]
    \centering
    \vspace{-0.1in}
    \resizebox{\columnwidth}{!}{%
    \includegraphics{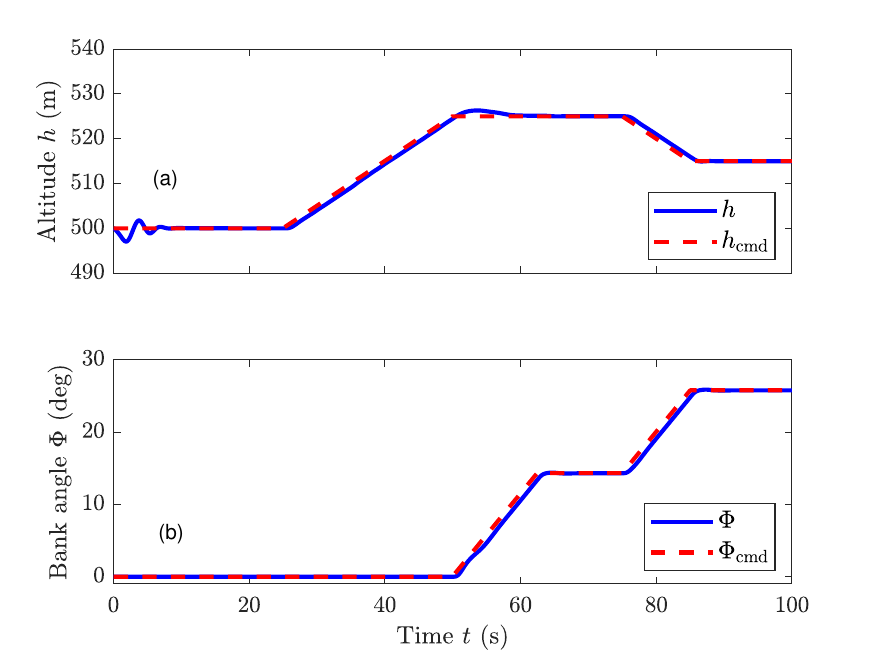}
    }
    \vspace{-0.2in}
    \caption{Altitude and bank-angle command following. (a) compares the command $h_{\rm cmd}$ and the simulated $h$. (b) compares the command $\Phi_{\rm cmd}$ and the simulated $\Phi$. Note that learning takes place for the longitudinal dynamics for $t\in[0,5]$ s causing the excitation seen in (a).}
    \vspace{-0.1in}
    \label{Case1Out}
\end{figure}

PCAC/long is initialized with $\psi_{\rm long,0} = 10^5 I_{36}$, $\theta_{\rm long,0} = 0.01 \textbf{1}_{36\times 1}$, $\hat{n} = 6$, $\ell = 40$, $\bar{Q} = 0.1 I_{39}$, $\bar{P} = 0.1$, $R = 0.1$, 
$\vert \Delta \delta e \vert \le \Delta \delta e_{\rm max} = 0.5$ deg/step, and $\vert \delta e \vert \le \delta e_{\rm max} = 10$ deg with no forgetting.
%
%
PCAC/lat is initialized with $\psi_{\rm lat,0} = 10^5 I_{36}$, $\theta_{\rm lat,0} = 0.01 \textbf{1}_{36\times 1}$, $\hat{n} = 6$, $\ell = 40$, $\bar{Q} = 0.1 I_{39}$, $\bar{P} = 0.1$, $R = 0.01$,
$\vert \Delta \delta a \vert \le \Delta \delta a_{\rm max} = 0.5$ deg/step, and $\vert \delta a \vert \le \delta a_{\rm max} = 2$ deg with no forgetting.
%
%
The resulting flight path data is shown in Figure \ref{Case1Out}.
An altitude ramp command is given for $t \in [25,50]$ s, followed by a bank-angle ramp command for $t\in[50,60]$.
Furthermore, altitude and bank-angle ramp commands are given simultaneously for $t \in [75,85]$ s.
As shown in Figure \ref{Case1Out}, PCAC follows the altitude and bank-angle commands with a slight overshoot in the altitude command following.

\begin{figure} [h!]
    \centering
    \vspace{-0.1in}
    \resizebox{\columnwidth}{!}{%
    \includegraphics{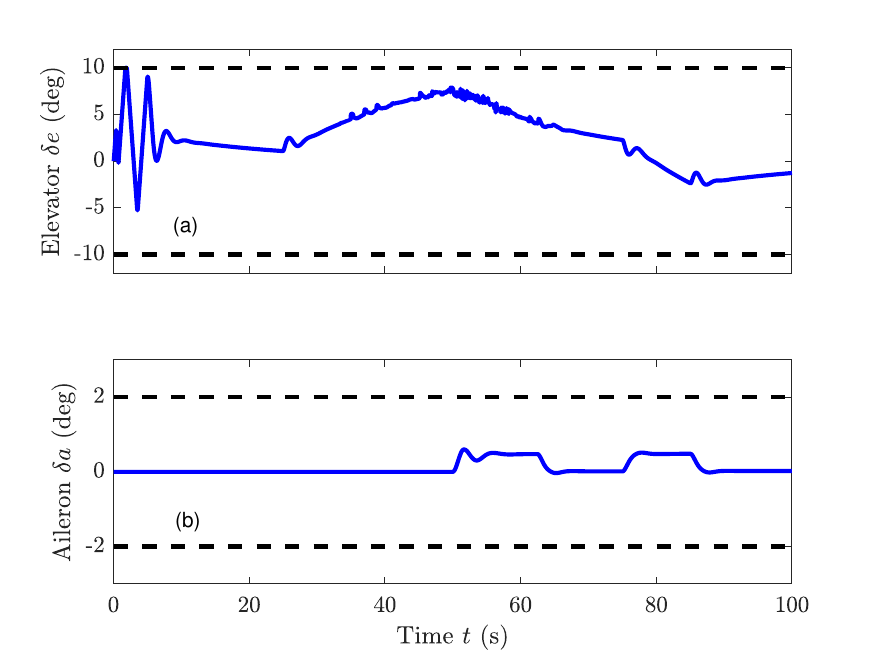}
    }
    \caption{Aileron and elevator deflections for altitude and bank-angle command following. (a) shows the elevator deflection $\delta e$ (blue) with the  constraints on $\vert \delta e \vert$ (dashed black). (b) shows the aileron deflection $\delta a$ (blue) with the constraints on $\vert \delta a \vert$  (dashed black).}
    \label{Case1Cont}
\end{figure}

\begin{figure} [h!]
    \centering
    \vspace{-0.1in}
    \resizebox{\columnwidth}{!}{%
    \includegraphics{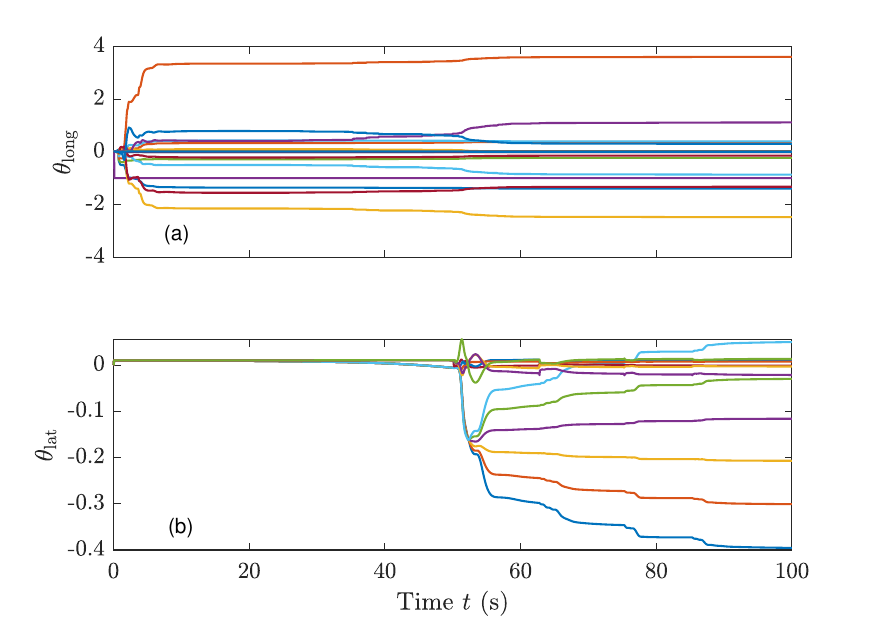}
    }
    \vspace{-0.2in}
    \caption{Estimated coefficient vectors for the altitude command-following loop $\theta_{\rm long}$ (a) and the bank-angle command-following loop $\theta_{\rm lat}$ (b). No prior modeling information is assumed, where  $\theta_{\rm long,0} = \theta_{\rm lat,0} = 0.01 \textbf{1}_{36\times 1}$.}
    \label{Case1ID}
    \vspace{-0.1in}
\end{figure}

Figures \ref{Case1Cont} and \ref{Case1ID} show the control-surface deflections $\delta e$ and $\delta a$ as well as the estimated coefficient vectors $\theta_{\rm long} $ and $\theta_{\rm lat}$, where $\theta_{\rm long}$ is the estimated coefficient vector of PCAC/long and $\theta_{\rm lat}$ is the estimated coefficient vector of PCAC/lat.
\hfill {\LARGE$\diamond$}

\subsection{Altitude and Azimuth-Angle Command Following} \label{AAT}

\begin{figure} [h!]
    \centering
    \resizebox{\columnwidth}{!}{%
    \includegraphics{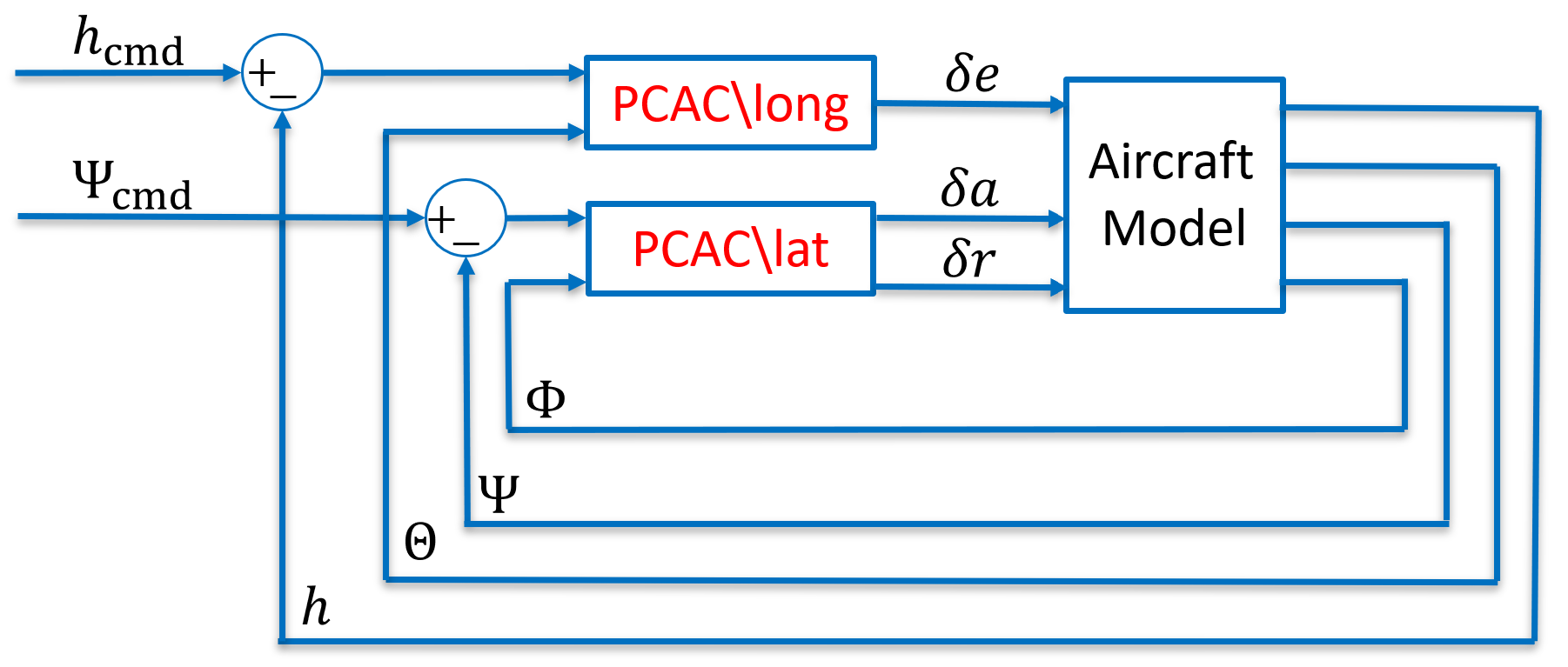}
    }
    \caption{Command-following block diagram for altitude $h$ and azimuth angle $\Psi$. 
    This architecture uses two PCAC controllers, namely, PCAC/long for the longitudinal dynamics and PCAC/lat for the lateral dynamics.
    Note that PCAC/long uses the elevation angle $\Theta$ for system identification and PCAC/lat uses the bank angle $\Phi$ for system identification.
    %
    }
    \vspace{-0.1in}
    \label{Case2BD}
\end{figure}

For this example, two PCAC loops are used to follow altitude and azimuth-angle commands.
The longitudinal loop uses the altitude $h$ and elevation angle $\Theta$ to identify the longitudinal dynamics and to specify the elevator deflection $\delta e$.
The lateral loop uses the bank angle $\Phi$ and azimuth angle $\Psi$ to identify the lateral dynamics and to specify the aileron and rudder deflections $\delta a$ and $\delta r$.
This control architecture is illustrated in Figure \ref{Case2BD}.

\begin{figure} [h!]
    \centering
    \vspace{-0.1in}
    \resizebox{\columnwidth}{!}{%
    \includegraphics{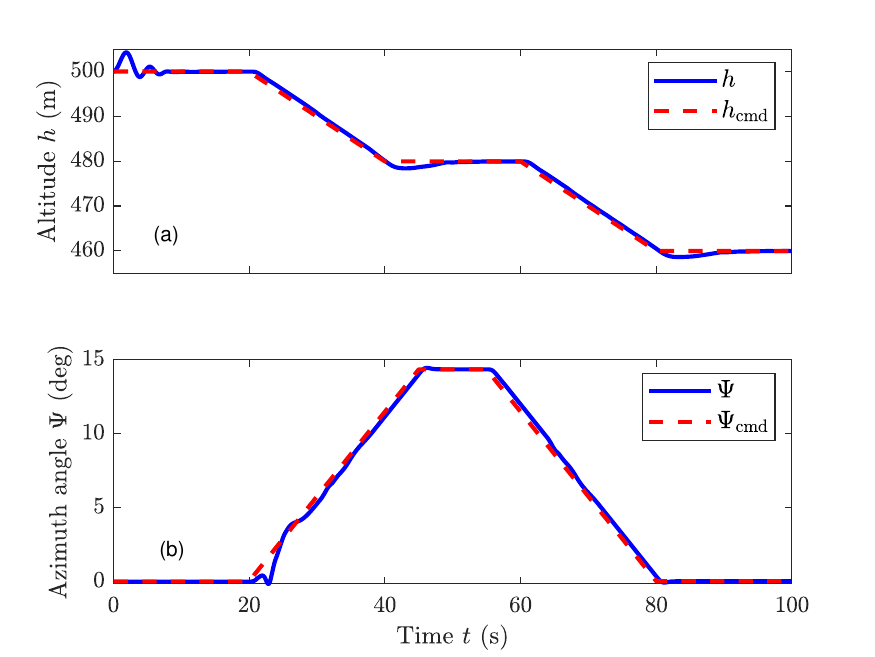}
    }
    \vspace{-0.1in}
    \caption{Altitude and azimuth-angle command following. (a) compares the command $h_{\rm cmd}$ and the simulated $h$. (b) compares the command $\Psi_{\rm cmd}$ and the simulated $\Psi$. Note that learning takes place for the longitudinal dynamics for $t \in [0,5]$ s causing the excitation seen in (a), while learning for the lateral dynamics takes place for $t \in [20,25]$ s, causing the excitation seen in (b).}
    \label{Case2Out}
    \vspace{-0.1in}
\end{figure}

The same PCAC/long parameters from Section \ref{ABT} were used to initialize the PCAC/long loop in this example.
%
%
PCAC/lat is initialized with $\psi_{\rm lat,0} = 2\times10^7 I_{48}$, $\theta_{\rm lat,0} = 0.01 \textbf{1}_{48\times 1}$, $\hat{n} = 6$, $\ell = 50$, $\bar{Q} = 60 I_{49}$, $\bar{P} = 60$, $R = 1.5I_{2\times2}$, 
$ [ \vert \Delta \delta a \vert, \ \vert \Delta \delta r \vert ]^{\rm T} \le [\Delta \delta a_{\rm max}, \ \Delta \delta r_{\rm max}]^{\rm T} = [0.2, \ 0.2]^{\rm T}$ deg/step, and $[ \vert \delta a \vert, \ \vert \delta r \vert]^{\rm T} \le [ \delta a_{\rm max} , \  \delta r _{\rm max}  ]^{\rm T} = [5, \ 5]^{\rm T}$ deg with no forgetting.
%
%
The resulting flight path data is shown in Figure \ref{Case2Out}.
Altitude ramp commands are given for $t \in [20,40]$ s and $t \in [60,80]$ s while azimuth-angle ramp commands are given for $t \in [20,45]$ s and $t \in [55,80]$ s.
As shown in Figure \ref{Case2Out}, PCAC follows the altitude and azimuth-angle commands with slight overshoot in the altitude command following.

\begin{figure} [h!]
    \centering
    \vspace{-0.1in}
    \resizebox{\columnwidth}{!}{%
    \includegraphics{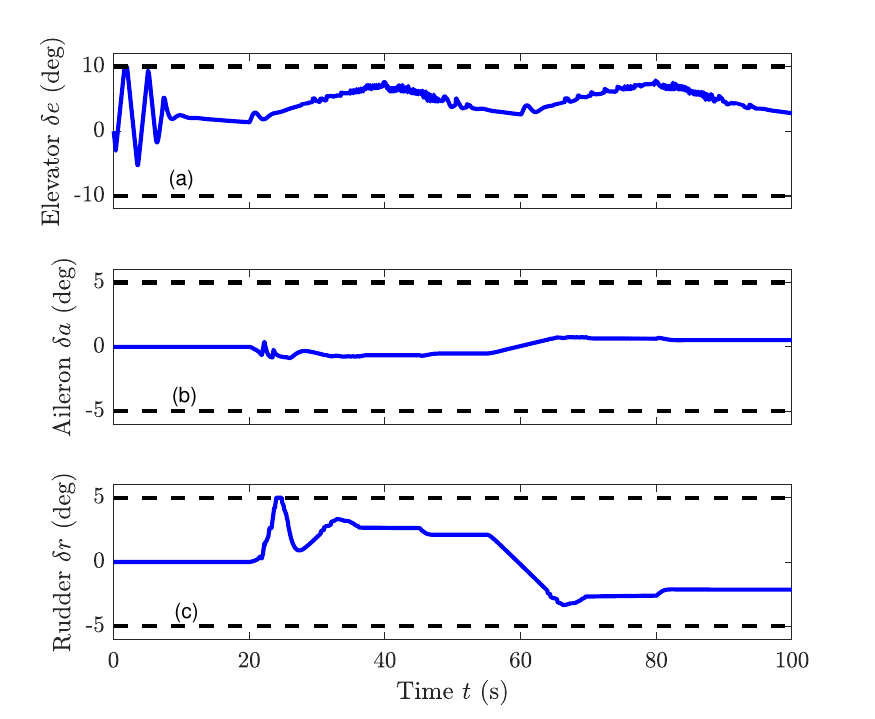}
    }
    \vspace{-0.1in}
    \caption{Elevator, aileron, and rudder deflections for altitude and azimuth-angle command following. (a) shows the elevator deflection $\delta e$ (blue) with the constraint on $\vert \delta e \vert$ (dashed black). (b) shows the aileron deflection $\delta a$ (blue) with the constraint on $\vert \delta a \vert$ (dashed black). (c) shows the rudder deflection $\delta r$ (blue) with the constraint $\vert \delta r \vert$ (dashed black).}
    \label{Case2Cont}
\end{figure}

\begin{figure} [h!]
    \centering
    \vspace{-0.1in}
    \resizebox{\columnwidth}{!}{%
    \includegraphics{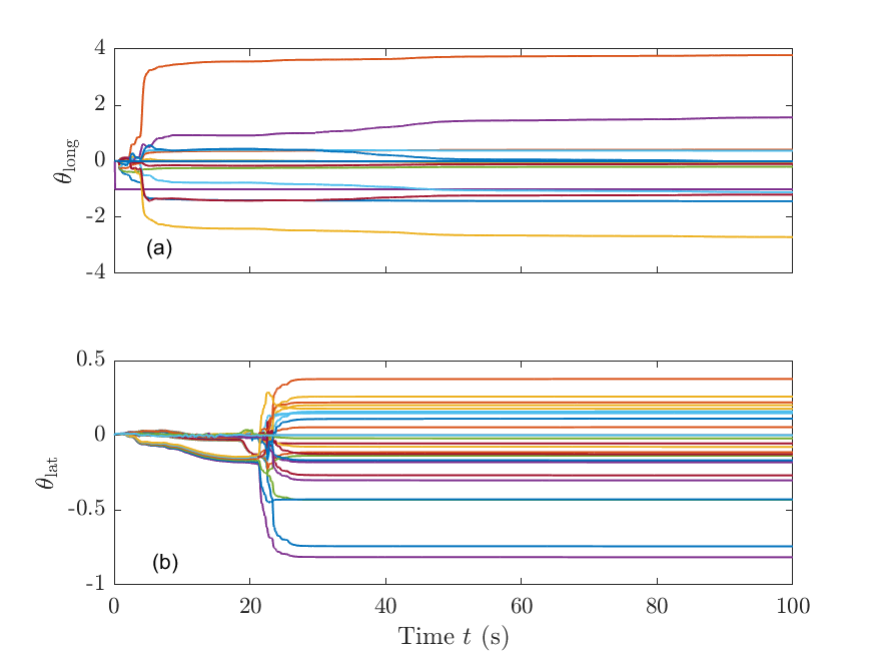}
    }
    \caption{Estimated coefficient vectors for the altitude command-following loop $\theta_{\rm long}$ (a) and the azimuth-angle command-following loop $\theta_{\rm lat}$ (b). No prior modeling information is assumed, where $\theta_{\rm long,0} = 0.01 \textbf{1}_{39\times1}$ and $\theta_{\rm lat,0} = 0.01 \textbf{1}_{49\times1}$.}
    \label{Case2ID}
    \vspace{-0.1in}
\end{figure}

Figures \ref{Case2Cont} and \ref{Case2ID} show the control surface deflections $\delta e$, $\delta a$, and $\delta r$ as well as the estimated coefficient vectors $\theta_{\rm long}$ and $ \theta_{\rm lat}$, where $\theta_{\rm long}$ is the estimated coefficient vector of PCAC/long and $\theta_{\rm lat}$ is the estimated coefficient vector of PCAC/lat.
\hfill {\LARGE$\diamond$}


\section{Adaptive Control of a Nonlinear 3DOF Aircraft Model} \label{sec:PCAC_exam2}

In this section, a nonlinear 3DOF fixed-wing passenger aircraft model provided by Matlab's aerospace toolbox 
and featured in \cite{aircraftSim}
is considered for longitudinal control.
The first example uses altitude and angle of attack measurements for altitude command following using the elevator. 
The second example uses altitude, angle of attack, and airspeed measurements for altitude and airspeed command following using the elevator and throttle.
Note that all simulations in this section use a sampling rate of $T_s = 0.05$ sec/step and the aircraft model is initialized at an altitude $h_0 = 2000$ m and airspeed $V_0 = 85$ m/s with all other initial model parameters set to 0. 
Both examples use forgetting for altitude command following with $\eta = 0.025$, $\tau_{\rm n} = 40$, $\tau_{\rm d} = 200$, and $\alpha_F = 0.001$.

 \subsection{Altitude Command Following} \label{AT}

\begin{figure} [h!]
    \centering
    \vspace{-0.1in}
    \resizebox{\columnwidth}{!}{%
    \includegraphics{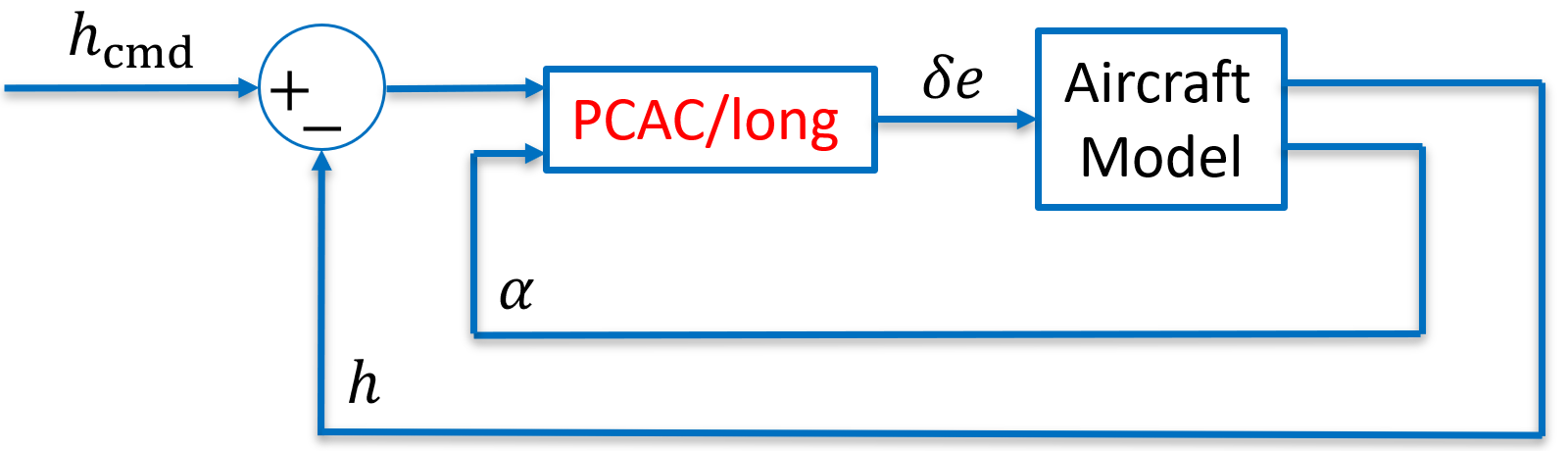}
    }
    \vspace{-0.1in}
    \caption{Command-following block diagram for altitude $h$. Note that the loop uses the angle of attack $\alpha$ for system identification.}
    \label{Case3BD}
    \vspace{-0.1in}
\end{figure}

 For this example, a single PCAC loop is used to follow altitude commands. 
 The loop uses the altitude $h$ and angle of attack $\alpha$ to identify the longitudinal dynamics and generate an elevator deflection $\delta e$. 
 This control architecture is illustrated in Figure \ref{Case3BD}.

\begin{figure} [h!]
    \vspace{-0.1in}
    \centering
    \resizebox{\columnwidth}{!}{%
    \includegraphics{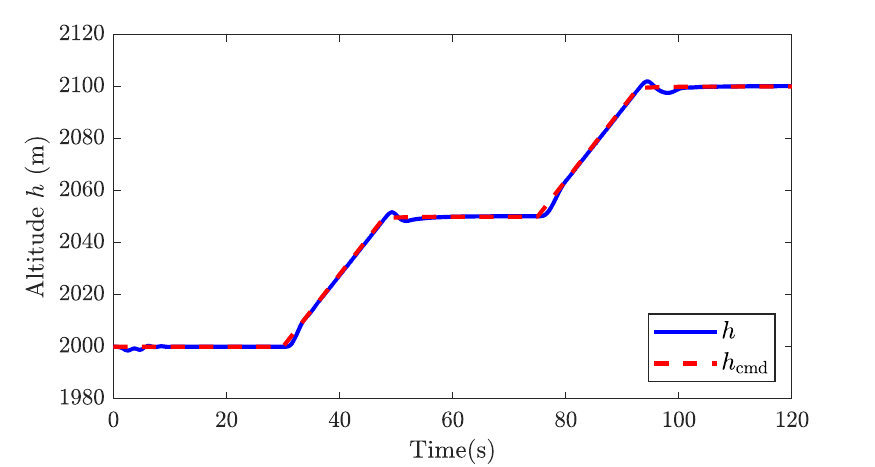}
    }
    \vspace{-0.1in}
    \caption{Altitude command following. The figure compares the command $h_{\rm cmd}$ and the simulated $h$. Note that learning takes place for $t \in [0,5]$ s causing the initial excitation seen in the plot.}
    \label{Case3Out}
    \vspace{-0.1in}
\end{figure}

PCAC/long is initialized with $\psi_{\rm long,0} = 10^4 I_{60}$, $\theta_{\rm long,0} = 0.01 \textbf{1}_{60\times 1}$, $\hat{n} = 10$, $\ell = 40$, $\bar{Q} = 0.01 I_{39}$, $\bar{P} = 500$, $R = 0.01$, 
$\vert \Delta \delta e \vert \le \Delta \delta e _{\rm max} = 1$ deg/step, and $\vert \delta e \vert \le \delta e_{\rm max} = 5$ deg.
%
%
The resulting flight path data is shown in Figure \ref{Case3Out}.
Altitude ramp commands are given for $t \in [30,50]$ s and $t \in [75,90]$ s.
As shown in Figure \ref{Case3Out}, PCAC follows the altitude commands with slight overshoot and undershoot.

\begin{figure} [h!]
    \centering
    \vspace{-0.1in}
    \resizebox{\columnwidth}{!}{%
    \includegraphics{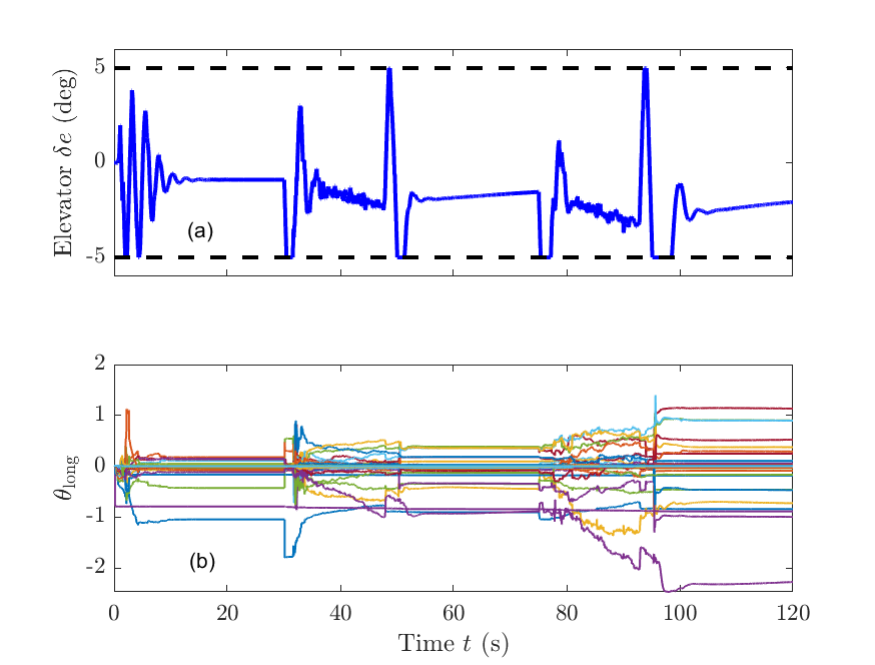}
    }
    \caption{Elevator deflection and estimated coefficient vector for altitude command following. (a) shows the elevator deflection $\delta e$ (blue) with the constraint on $\vert \delta e \vert$ (dashed black). (b) shows the estimated coefficient vector $\theta_{\rm long}$. No prior modeling information is assumed, where  $\theta_{\rm long,0} = 0.01 \textbf{1}_{60\times 1}$. }
    \label{Case3ContID}
    \vspace{-0.1in}
\end{figure}

Figure \ref{Case3ContID} shows the control surface deflection $\delta e$ and estimated coefficient vector $\theta_{\rm long}$.
\hfill {\LARGE$\diamond$}

\subsection{Altitude and Airspeed Command Following} \label{AVT}

\begin{figure} [h!]
    \centering
    \resizebox{\columnwidth}{!}{%
    \includegraphics{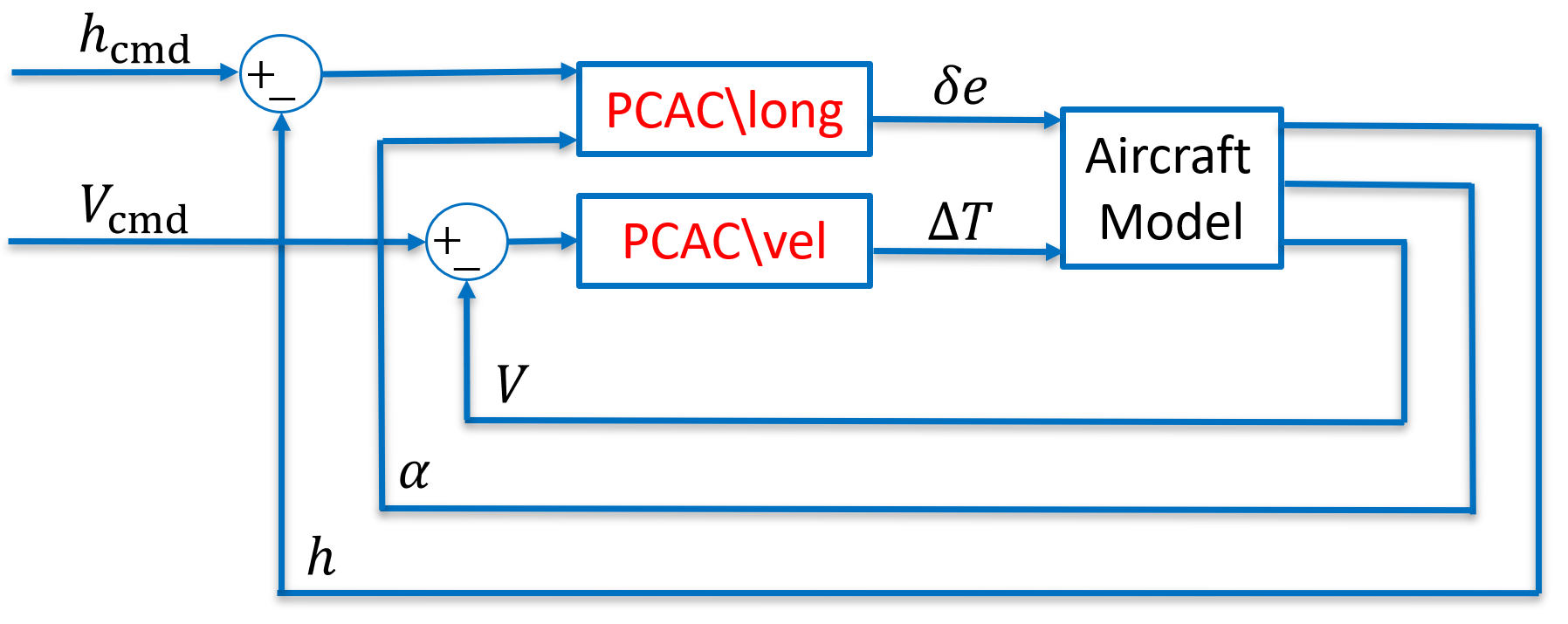}
    }
    \vspace{-0.1in}
    \caption{Command-following block diagram for altitude $h$ and airspeed $V$. This architecture uses two PCAC controllers, namely, PCAC/long for the longitudinal dynamics and PCAC/vel for the throttle dynamics. Note that PCAC/long uses the angle of attack $\alpha$ for system identification.}
    \label{Case4BD}
    \vspace{-0.1in}
\end{figure}

For this example, two PCAC loops are used to follow altitude and airspeed commands. 
The longitudinal loop uses the altitude $h$ and angle of attack $\alpha$ to identify the longitudinal dynamics and to specify the elevator deflection $\delta e$.
The airspeed loop uses the airspeed $V$ to identify the throttle dynamics and generate a change in throttle $\delta T$.
Note that the throttle parameter $T \in [0,1]$ has a constant input of 0.5.
This control architecture is illustrated in Figure \ref{Case4BD}. 

\begin{figure} [h!]
    \centering
    \vspace{-0.1in}
    \resizebox{\columnwidth}{!}{%
    \includegraphics{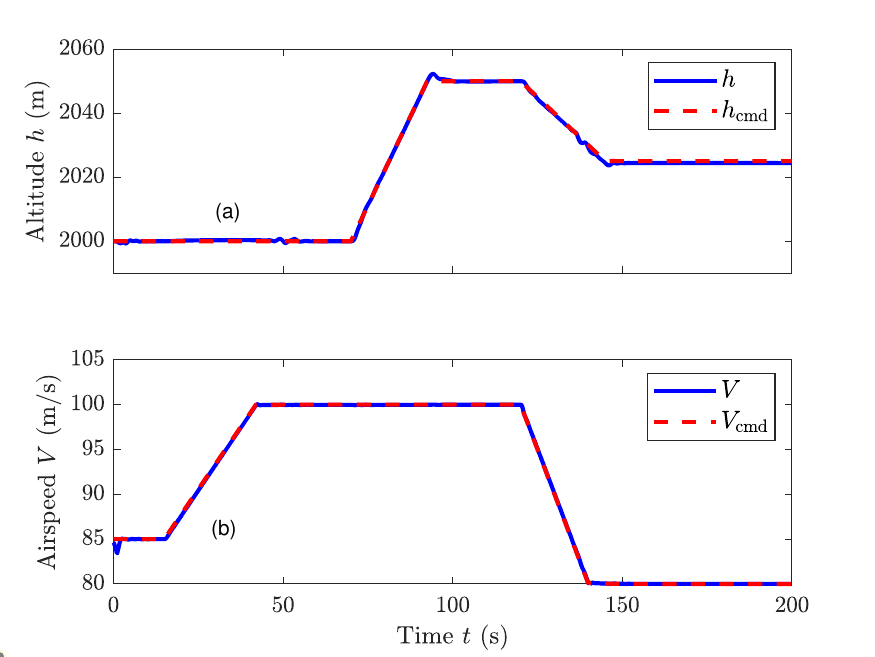}
    }
    \vspace{-0.1in}
    \caption{Altitude and airspeed command following. (a) compares the command $h_{\rm cmd}$ and the simulated $h$. (b) compares the command $V_{\rm cmd}$ and the simulated data $V$. Note that learning takes place for both the longitudinal dynamics and throttle dynamics for $t \in [0,5]$ s causing the initial excitations seen in the plots.}
    \label{Case4Out}
    \vspace{-0.1in}
\end{figure}

The same PCAC/long parameters from Section \ref{AT} were used to initialize the PCAC/long altitude command-following loop.
PCAC/vel is initialized with $\psi_{\rm vel,0} = 10^5 I_{8}$, $\theta_{\rm vel,0} = 0.01 \textbf{1}_{8\times 1}$, $\hat{n} = 4$, $\ell = 30$, $\bar{Q} = 0.0001 I_{29}$, $\bar{P} = 0.01$, $R = 0.01$,
$\vert \Delta \delta T \vert \le \Delta \delta T_{\rm max} = 0.1$ 1/step, and $\vert \delta T \vert \le \delta T_{\rm max} = 0.25$ with no forgetting.
%
%
The resulting flight path data is shown in Figure \ref{Case4Out}.
Airspeed ramp commands are given for $t \in [10,45]$ s followed by an altitude ramp command for $t \in [70,95]$ s. Furthermore, altitude and airspeed commands are given simultaneously for $t \in [125,145]$ s.
As shown in Figure \ref{Case4Out}, PCAC follows the altitude and airspeed commands with slight overshoot in the altitude command following. 

\begin{figure} [h!]
    \centering
    \vspace{-0.1in}
    \resizebox{\columnwidth}{!}{%
    \includegraphics{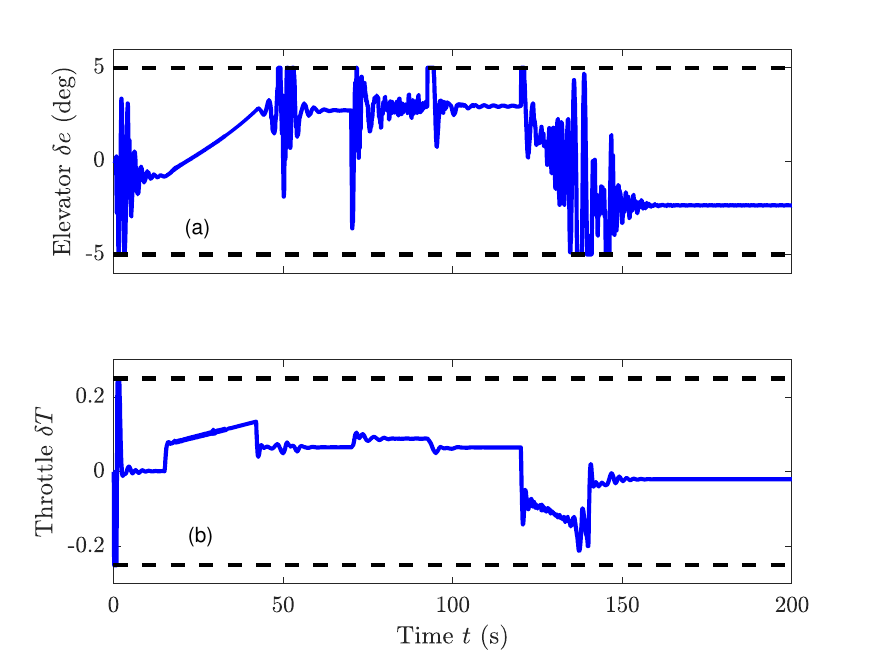}
    }
    \vspace{-0.1in}
    \caption{Elevator deflection and throttle change for altitude and airspeed command following. (a) shows the elevator deflection $\delta e$ (blue) with the constraint on $\vert \delta e \vert$ (dashed black). (b) shows the throttle change $\delta T$ (blue) with the constraint on $\vert \delta T \vert$ (dashed black).}
    \label{Case4Cont}
\end{figure}

\begin{figure} [h!]
    \centering
    \vspace{-0.1in}
    \resizebox{\columnwidth}{!}{%
    \includegraphics{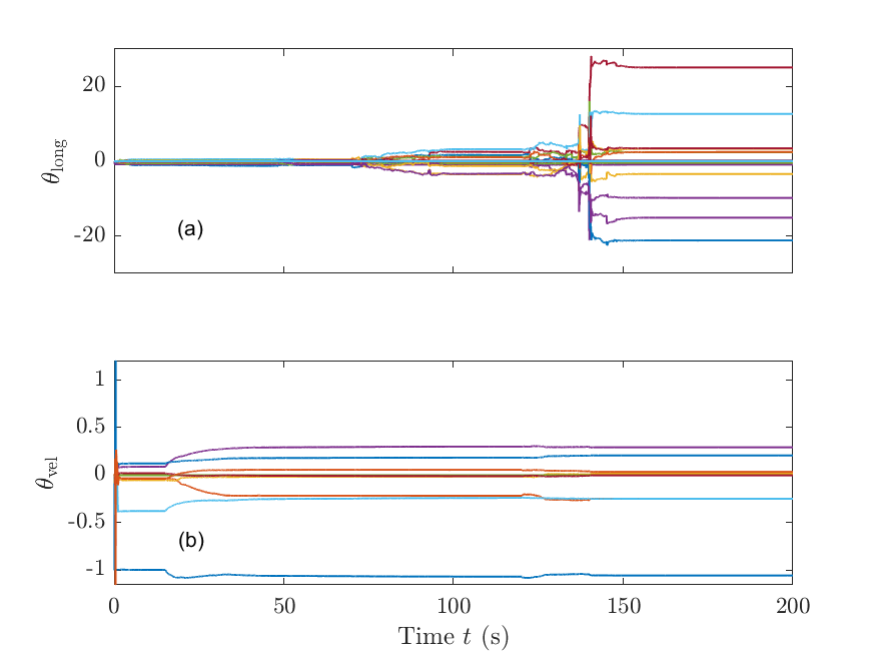}
    }
    \caption{Estimated coefficient vectors for the altitude command-following loop $\theta_{\rm long}$ (a) and the airspeed command-following loop $\theta_{\rm vel}$ (b). No prior modeling information is assumed, where $\theta_{\rm long,0} = 0.01 \textbf{1}_{60 \times 1}$ and $\theta_{\rm vel,0} = 0.01 \textbf{1}_{8 \times 1}$.}
    \label{Case4ID}
    \vspace{-0.1in}
\end{figure}

Figures \ref{Case4Cont} and \ref{Case4ID} show the control surface deflection $\delta e$, throttle changes $\delta T$, and estimated coefficient vectors $\theta_{\rm long}$ and $\theta_{\rm vel}$, where $\theta_{\rm long}$ is the estimated coefficient vector of PCAC/long and $\theta_{\rm vel}$ is the estimated coefficient vector of PCAC/vel.
\hfill {\LARGE$\diamond$}


\section{Conclusions and future work} \label{sec:conclusions}

For autonomous control of aircraft with unmodeled aerodynamics, this paper proposed an autopilot based on predictive cost adaptive control (PCAC), which is an indirect adaptive control extension of model predictive control.
PCAC uses recursive least squares (RLS) with variable-rate forgetting for online, closed-loop system identification, and receding-horizon optimization based on quadratic programming.
This technique was demonstrated numerically on a 6DOF linearized aircraft model and a 3DOF nonlinear aircraft model.
In both cases, the adaptive autopilot was able to follow attitude, bank-angle, azimuth-angle, and velocity commands over a range of operation.

The ability of PCAC to operate as an adaptive autopilot without aerodynamic modeling has useful implications in practice.
First and foremost, this technique can mitigate the need for extensive wind tunnel testing.
Second, it can help avoid the need for gain scheduling and dynamic inversion.
Finally, PCAC can be used to accelerate the expensive and time-consuming aircraft/autopilot design cycle.
This potential will be investigated in future research through embedded control of fixed-wing, autonomous aircraft.



\bibliographystyle{IEEEtran}
\bibliography{IEEEabrv,bib_paper.bib,bib2.bib,bib3.bib}

\end{document}